# Ferroelectricity induced by ferriaxial crystal rotation and spin helicity in a $B$-site-ordered double-perovskite multiferroic In$_2$NiMnO$_6$

Noriki Terada,[1,*] Dmitry D. Khalyavin,[2] Pascal Manuel,[2] Wei Yi,[3] Hiroyuki S. Suzuki,[1]
Naohito Tsujii,[1] Yasutaka Imanaka,[1] and Alexei A. Belik[3,†]

[1]*National Institute for Materials Science, Sengen 1-2-1, Tsukuba, Ibaraki 305-0047, Japan*
[2]*ISIS Facility, STFC Rutherford Appleton Laboratory, Chilton, Didcot, Oxfordshire, OX11 0QX, United Kingdom*
[3]*International Center for Materials Nanoarchitectonics (WPI-MANA), National Institute for Materials Science, 1-1 Namiki, Tsukuba, Ibaraki 305-0044, Japan*

We have performed dielectric measurements and neutron diffraction experiments on the double perovskite In$_2$NiMnO$_6$. A ferroelectric polarization, $P \simeq 30\ \mu$C/m$^2$, is observed in a polycrystalline sample below $T_N = 26$ K where a magnetic phase transition occurs. The neutron diffraction experiment demonstrates that a complex noncollinear magnetic structure with "cycloidal" and "proper screw" components appears below $T_N$, which has the incommensurate propagation vector $\bm{k} = (k_a, 0, k_c; k_a \simeq 0.274, k_c \simeq -0.0893)$. The established magnetic point group $21'$ implies that the macroscopic ferroelectric polarization is along the monoclinic $b$ axis. Recent theories based on the inverse Dzyaloshinskii-Moriya effect allow us to specify two distinct contributions to the polarization of In$_2$NiMnO$_6$. One of them is associated with the cycloidal component, $\bm{p}_1 \propto \bm{r}_{ij} \times (\bm{S}_i \times \bm{S}_j)_\perp$, and the other with the proper screw component, $\bm{p}_2 \propto [\bm{r}_{ij} \cdot (\bm{S}_i \times \bm{S}_j)_\parallel]\bm{A}$. The latter is explained by coupling between spin helicity and "ferriaxial" crystal rotation with macroscopic ferroaxial vector $\bm{A}$, characteristic of the $B$-site ordered perovskite systems with out-of-plane octahedral tilting.

## I. INTRODUCTION

Magnetoelectric multiferroic compounds, which have ferroelectric and (anti)ferromagnetic orderings, have attracted much attention in the past decade [1,2]. In particular, since the discovery of multiferroicity in the perovskite TbMnO$_3$ [3], many other rare earth perovskite manganites, $R$MnO$_3$ ($R$ = rare earth or Y), have been intensively studied. In some cases, such as TbMnO$_3$ and DyMnO$_3$, the magnetic orderings are cycloidal structures, inducing a ferroelectric polarization through an inverse Dzyaloshinskii-Moriya (DM) effect [4–8]. The other type of noncollinear magnetic ordering, proper screw helical, also induces a ferroelectric polarization in crystals with a "ferroaxial" point group. In this case, the polarization is explained as a coupling between the spin chirality, $\bm{r}_{ij} \cdot (\bm{S}_i \times \bm{S}_j)$, and ferroaxial crystal rotation, $\bm{A}$, based on the inverse DM effect [9–11]. While the orthorhombic perovskites with $Pbnm$ space group are not in the ferroaxial group, $B$-site ordered perovskites with the monoclinic $P2_1/n$ belong to ferroaxial class, as illustrated in Fig. 1.

A number of $B$-site ordered double-perovskite materials with different cation compositions have been studied since the 1960s [12–14]. Recently, for double-perovskites oxides, $R_2B'B''O_6$, emergence of multiferroic behavior has been predicted by theoretical calculation for Y$_2$NiMnO$_6$ [15] and has been experimentally reported for Lu$_2$MnCoO$_6$ [16] and Y$_2$MnCoO$_6$ [17]. Although the other rare earth double-perovskites consisting of Ni and Mn are simple ferromagnets [18,19], the theory predicted the $E$ type, $\uparrow\uparrow\downarrow\downarrow$, antiferromagnetic ordering in Y$_2$NiMnO$_6$ [15]. Relationships between magnetic orderings and ferroelectric polarization in these compounds, however, have not been fully understood thus far.

More recently, Yi et al. have succeeded in synthesizing a new double-perovskite compound In$_2$NiMnO$_6$ (INMO) by means of high-pressure and high-temperature technique [20,21]. The crystal structure of the monoclinic space group $P2_1/n$ is shown in Fig. 1(a). One of the $B$ sites (2c) is occupied by Mn$^{4+}$ with the electronic configuration ($d^3$: $t_{2g}^3 e_g^0$), while Ni$^{2+}$ with $3d^8$ : $t_{2g}^6 e_g^2$ is placed on the other $B$ site (2d). An antiferromagnetic phase transition at $T_N = 26$ K was concluded based on magnetic susceptibility and specific heat measurements. In the present work, in order to investigate the magnetic ordering and existence of ferroelectricity, we have performed a neutron diffraction experiment and dielectric constant and polarization measurements on polycrystalline samples of INMO. Our study reveals a ferroelectric polarization induced by a combination of spin helicity with $B$-site chemical ordering and octahedral tilting resulting in a "ferriaxial" mechanism unique for the ordered double perovskites.

## II. EXPERIMENTAL DETAILS

Powder samples of INMO were prepared by solid-state reaction method under high pressure at the National Institute for Materials Science (NIMS) [20,21]. Small amounts of impurity phases, In$_2$Mn$_2$O$_7$ and NiO, were found in our neutron diffraction experiment. The former has a ferromagnetic phase transition at 121 K [22]. The latter is known as a simple antiferromagnet with the Néel temperature above room temperature [23]. Therefore, these impurity phases should not affect the physical properties of the In$_2$NiMnO$_6$ main phase below 26 K.

*terada.noriki@nims.go.jp
†alexei.belik@nims.go.jp

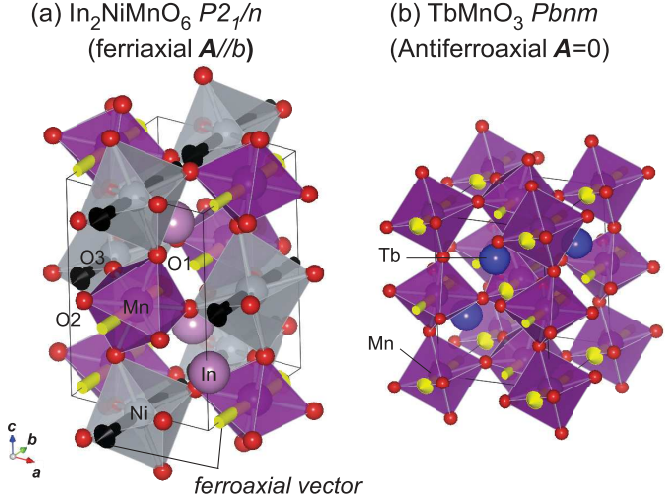

FIG. 1. (Color online) (a) Double-perovskite crystal structure of In$_2$NiMnO$_6$. The space group is the monoclinic $P2_1/n$ (no. 14: cell choice 2). The lattice constants are $a = 5.13520(1)$ Å, $b = 5.33728(1)$ Å, $c = 7.54559(4)$ Å, and $\beta = 90.1343(1)°$ [20,21]. (b) Crystal structure of the orthorhombic perovskite TbMnO$_3$ with $Pbnm$ space group. The vectors denote local ferroaxial vectors corresponding to tilting of oxygen octahedra from a cubic perovskite.

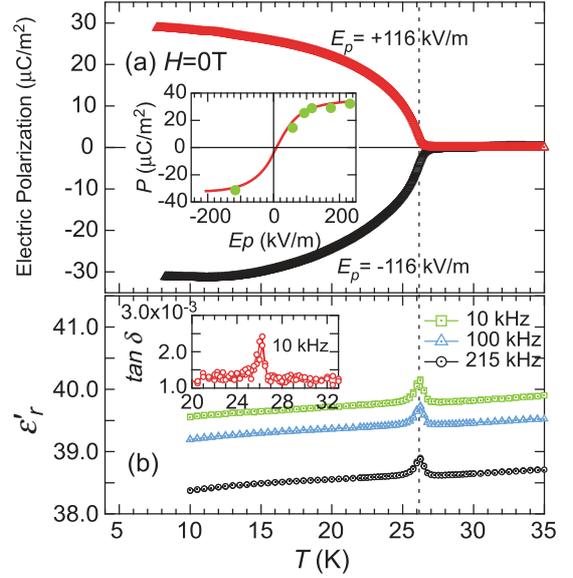

FIG. 2. (Color online) Temperature dependence of (a) electric polarization and (b) real part of the dielectric constant, and the loss tangent in the inset for a powder sample of In$_2$NiMnO$_6$. The inset in panel (a) is the poling electric field dependence of the polarization.

The dielectric constant and pyroelectric current measurements were done by Agilent E4980A LCR meter and Keithley 6517B electrometer combined with the physical property measurement system of quantum design. The dielectric properties were measured using hardened pellets (9.0 mm$^2$ covered with sliver paste × 0.8 mm thickness) of the polycrystalline INMO. The magnetization measurements were carried out by an extraction method with a 15-T superconducting magnet in NIMS. The neutron powder diffraction measurements were carried out on the cold neutron time-of-flight diffractometers WISH [24] at the ISIS Facility of the Rutherford Appleton Laboratory (UK). We measured 300 mg INMO powder for this experiment. We used a He cryostat to cool the sample down to 1.5 K. The crystal and magnetic structure refinements were performed using the FullProf program [25].

## III. EXPERIMENTAL RESULTS

As shown in Fig. 2(a), electric polarization appears below $T_N = 26$ K. The polarization can be switched by reversing the poling electric field ($E_p$), indicating emergence of ferroelectricity below $T_N$. The ferroelectric polarization almost reaches a saturation value, $P \simeq 30 \, \mu$C/m$^2$, above $E_p = 100$ kV/m, which is shown in the inset of Fig. 2(a). Since a powder sample was used in the measurements, $P$ is a half of the intrinsic value for a single crystal, namely $P_{\text{intrinsic}} \simeq 60 \, \mu$C/m$^2$, which is comparable to other spin-driven ferroelectrics [2]. The real part of dielectric constant, $\epsilon'$, also shows a peak anomaly at $T_N$ in all frequencies measured, which is in agreement with the ferroelectric phase transition. Energy dissipation also displays an anomaly at the same temperature [Fig. 2(b)].

With decreasing temperature from 40 K in the paramagnetic phase, a set of magnetic reflections was observed below $T_N = 26$ K, as shown in Figs. 3 and 4(a). As seen in the inset of Fig. 3, a magnetic reflection was also observed in the low-$Q$ region around $Q \simeq 0.34$ Å$^{-1}$. All the magnetic reflections observed below $T_N$ can be indexed by the incommensurate wave vector, $\boldsymbol{k} = (k_a, 0, k_c)$, e.g., $(0.274, 0, -0.0893)$, at 1.5 K. The propagation vector depends on temperature for 15 K $\leqslant T \leqslant 26$ K, as shown in Fig. 4(b). The line width of these reflections is resolution limited, indicating that the spin correlation length of the long period magnetic ordering is infinite at low temperature. These experimental data drive us to the conclusion that the incommensurate magnetic ordering is closely related to the ferroelectric polarization that appears below $T_N$.

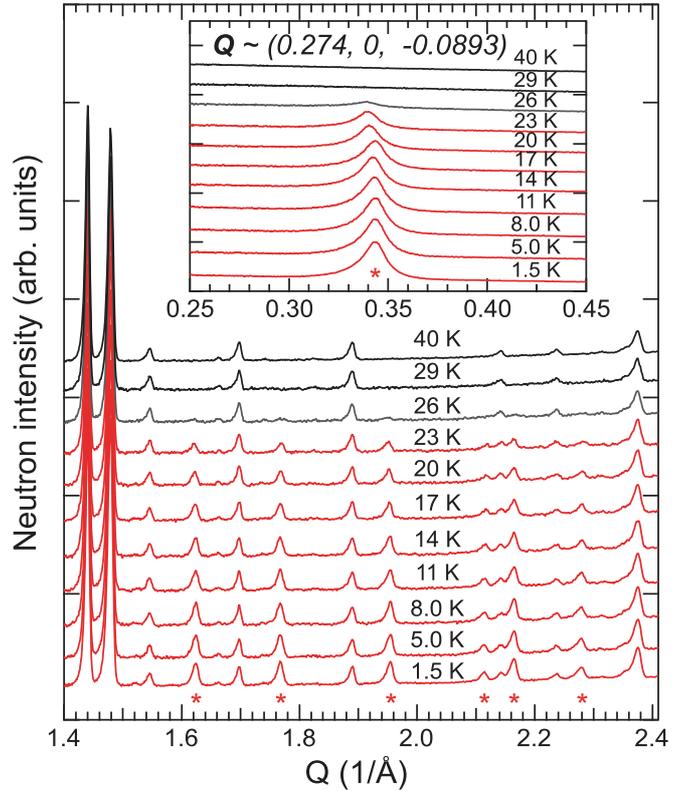

FIG. 3. (Color online) Temperature dependence of the neutron diffraction profile obtained by using 90-deg detector banks on WISH. The inset shows the magnetic reflection in low-$Q$ region, which was observed in forward scattering. Stars indicate the positions of magnetic reflection.

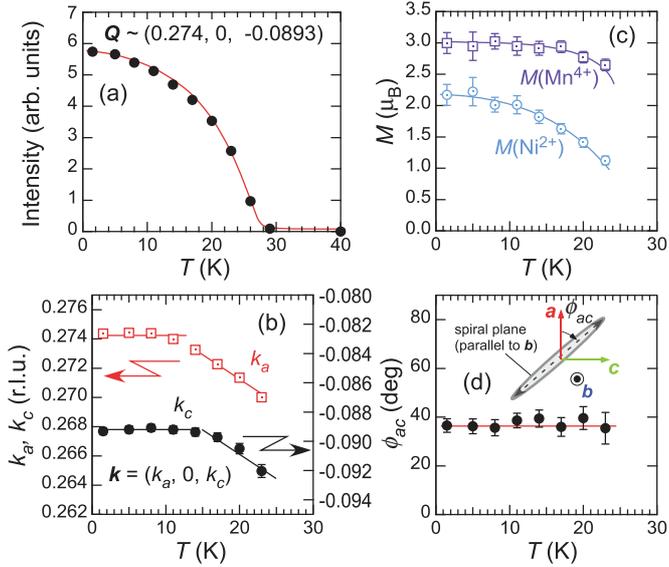
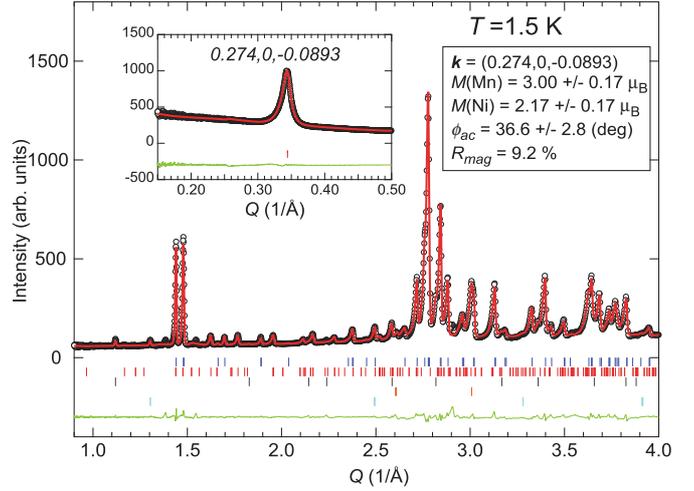

FIG. 4. (Color online) Temperature dependence of (a) integrated intensity of the magnetic reflection at $(0.274, 0, -0.0893)$, (b) $a$ and $c$ components of the incommensurate magnetic propagation wave vector, (c) magnetic moments of $Mn^{4+}$ and $Ni^{2+}$ spins, and the angle between $a$ axis and the spiral plane, including $b$ axis, in $ac$ plane.

FIG. 5. (Color online) Typical result of the magnetic structural refinement for the experimental data at 1.5 K. The first and second lines of the vertical bars indicate the peak positions for the crystallographic and magnetic phases of $In_2NiMnO_6$. The third, fourth, and fifth bars denote impurity phases of $In_2Mn_2O_7$, NiO, and magnetic phase of NiO. The refined parameters and reliability factors for the data at 1.5 K are written in the right inset.

As shown in Fig. 5, we successfully refined the diffraction data at 1.5 K, using the $P2_1/n$ space group for the nuclear scattering and the noncollinear spin structure (Fig. 6) for the magnetic reflections. The reliability factor was $R_{mag} = 9.2\%$. We confirmed from the neutron powder diffraction that Ni and Mn atoms are fully ordered. The noncollinear magnetic structure has two orthogonal spin components along the $b$ axis and in the $ac$ plane. The latter spin component is tilted by $\phi_{ac} = 36.6°$ from the $a$ axis as illustrated in Fig. 6. The refined magnetic moments of $Mn^{4+}$ and $Ni^{2+}$ are $M(Mn) = 3.00 \pm 0.17 \mu_B$ and $M(Ni) = 2.17 \pm 0.17 \mu_B$ at 1.5 K, which are consistent with theoretical values of $Mn^{4+}$ ($S = 3/2$) and $Ni^{2+}$ ($S = 1$). The temperature dependences of $M(Mn)$ and $M(Ni)$ are shown in Fig. 4(c). The tilting angle $\phi_{ac}$ is independent of temperature, which is shown in Fig. 4(d). While the spin ordering along the $b$ axis is ferromagnetic coupling due to zero value of the $b$ component in $\boldsymbol{k}$, those along both $a$ and $c$ directions are spiral orders (Fig. 6). Since the $\boldsymbol{k}$ vector is neither parallel nor perpendicular to the spiral plane, this noncollinear magnetic ordering has both "proper screw" and "cycloid" components.

The magnetic order parameter is expressed as the superposition of two time-odd irreducible representations of the $P2_1/n$

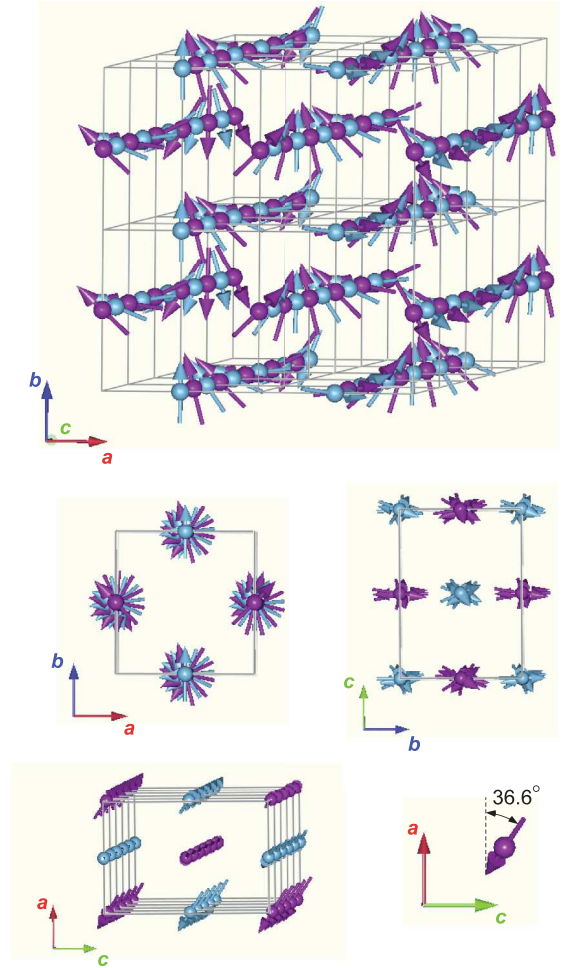

FIG. 6. (Color online) Illustrations of the noncollinear magnetic structure of $In_2NiMnO_6$.

space group, $mF_1 \oplus mF_2$, combined with real and imaginary characters [26]. The two representations transform the spin-density waves (with the spin components being along the $b$ axis and within the $ac$ plane) which are degenerate in respect of the exchange energy and therefore belong to the same exchange multiplet. Thus, the found magnetic structure corresponds to a single irreducible representation of the exchange Hamiltonian. Magnetocrystalline anisotropy splits the components of the exchange multiplet, imposing a first-order character for the transition with the reducible order parameter. In the case of In$_2$MnNiO$_6$, there is no clear evidence of discontinuity in our neuron data and the previously reported specific heat data [20], implying that the anisotropic interactions are small and the transition is only weakly first order. An alternative explanation of the reducible nature of the magnetic order parameter can be a sequence of two transitions with very close critical temperatures beyond the resolution of our experimental data. The noncollinear magnetic ordering breaks the symmetry elements of $P2_1/n$, $n$-glide perpendicular to $b$ and inversion, and leaves a two-fold screw along $b$. The magnetic point group is the polar $21'$ that allows the ferroelectric polarization along the $b$ axis.

As seen in Figs. 7(a) and 7(b), the ferroelectric polarization is preserved up to ∼5 T, while it disappears above 5 T. As reported in a previous paper [20,21], the phase transition temperature goes down with increasing magnetic field. A spin-flop-like anomaly was observed at $H_{sf} = 1.8$ T, which is clearly seen in the $dM/dH$ curve in Fig. 7(c). However, no anomaly was found around $H_{sf}$ in the electric polarization [Fig. 7(b)]. The field, 5 T, where the ferroelectric polarization disappears corresponds to phase transition field to a ferromagnetic phase, namely $H_{st}$. Since the magnetic phase below $H_{sf}$ has been identified as the complex noncollinear spin state, another polar magnetic state should be realized for $H_{sf} \leqslant H \leqslant H_{st}$. For further understanding of the field-induced phase, a single crystal experiment under magnetic field is needed.

## IV. DISCUSSION

Let us discuss the microscopic mechanism regarding the emergence of ferroelectric polarization of INMO, based on the determined magnetic structure and previous theoretical works. The $21'$ symmetry restricts the polarization along the $b$ axis. For the determined noncollinear magnetic structure in INMO, the spiral plane restricts neither parallel nor perpendicular to the $k$ vector. We can decompose the vector, $S_i \times S_j$, into two components perpendicular and parallel to $k$, namely $(S_i \times S_j)_\perp$ and $(S_i \times S_j)_\parallel$, respectively [Fig. 8(a)]. Considering the well-known theory for cycloid ordering based on inverse DM effect [4,5] or spin current mechanism [6], represented by $p \propto r_{ij} \times (S_i \times S_j)_\perp (\equiv p_1)$, the cycloidal component shown in Fig. 8(b) yields an electric polarization perpendicular to both $k$ and $(S_i \times S_j)_\perp$, namely along the $b$ axis. As mentioned at the beginning, the crystal structure of INMO belongs to the ferroaxial point group $2/m$. If the crystal structure was of $Pbnm$ without Mn and Ni ordering, the proper screw structure would be nonpolar ($2221'$). In this $B$-site-ordered case, however, there exists a macroscopic nonzero ferroaxial vector, due to the difference in the degree of oxygen octahedral tilting between Mn and Ni sites. Although these local ferroaxial vectors have opposite directions on the neighbor sites, their absolute values are different (unlike the case of the "antiferroaxial" $Pbnm$ structure of $TbMnO_3$), resulting in "ferriaxial" crystal ordering [Fig. 1(b)]. In this case, the proper screw component, where the $(S_i \times S_j)_\parallel$ is parallel to $k$, induces the other ferroelectric polarization component along the $b$ axis, $p_2 \propto [r_{ij} \cdot (S_i \times S_j)_\parallel]A$ [Fig. 8(c)]. because the product $p_2 \cdot [r_{ij} \cdot (S_i \times S_j)_\parallel]A$ is invariant under all symmetry operations of $P2_1/n1'$.

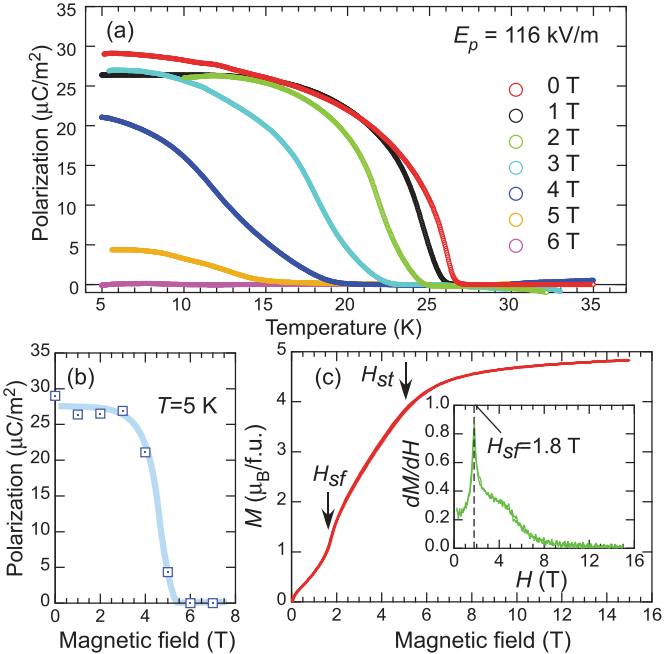

FIG. 7. (Color online) (a) Temperature dependence of the electric polarization in applied magnetic fields up to 6 T. Magnetic field dependence of (b) the electric polarization at 5 K up to 7 T and (c) magnetization at 1.6 K up to 15 T. The inset in panel (c) is derivative of the magnetization with respect to magnetic field.

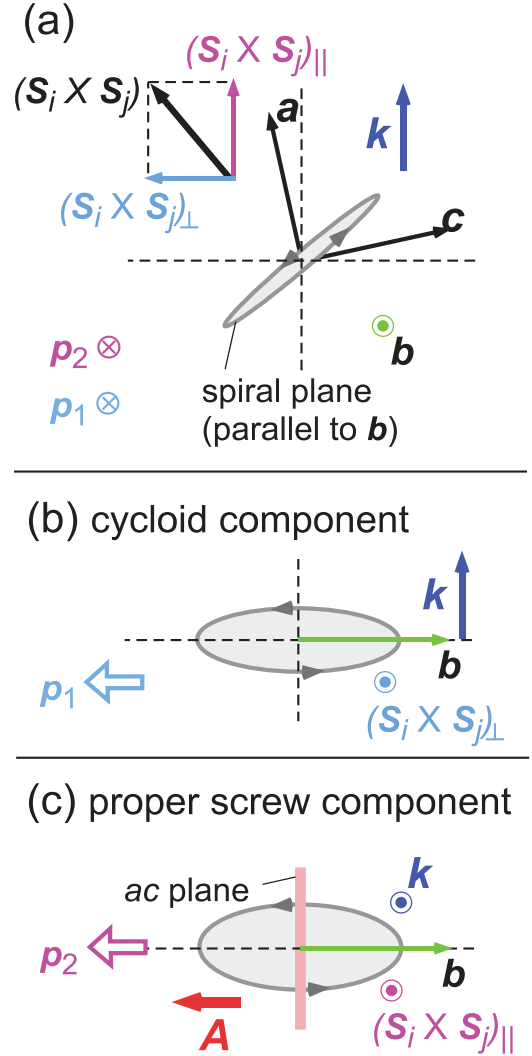

FIG. 8. (Color online) (a) Schematic drawing of the noncollinear structure of INMO, denoting the relationship among spin helicity vector, propagation vector $k$, and polarization. Panels (b) and (c) show the projections from directions perpendicular and parallel to $k$.

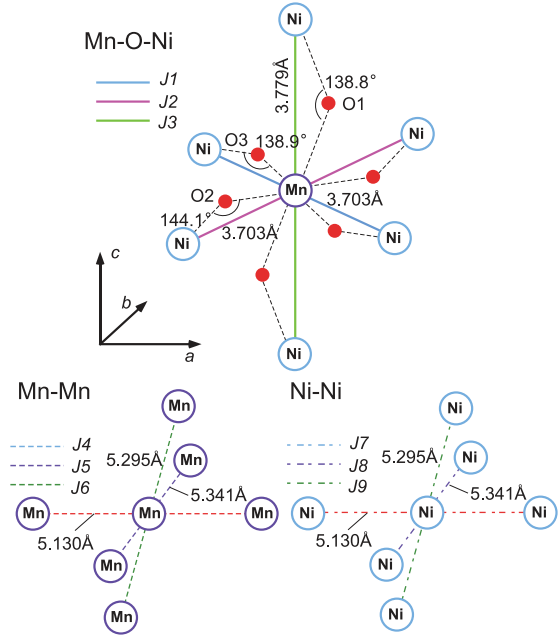

FIG. 9. (Color online) Schematic drawings of neighboring exchange bonds connecting $Mn^{4+}$ and $Ni^{2+}$, $Mn^{4+}$ and $Mn^{4+}$, and $Ni^{2+}$ and $Ni^{2+}$ in $In_2NiMnO_6$. The bond angles and distances were taken from the neutron diffraction data at 40 K.

The noncollinear magnetic structure determined in INMO is rare in perovskite compounds. Although some rare earth manganites, such as $RMnO_3$ ($R$ = Gd, Tb, Dy), have cycloidal magnetic structures, their magnetic propagation vectors are along the orthorhombic $b$ axis, $k = (0, q, 0)$ [1,2]. Other rare earth manganites with collinear structure possess $k$ along the $b$ axis as well [27–29]. The rich magnetic phase diagram as functions of ionic radius and temperature in orthorhombic $RMnO_3$ is caused by strong spin frustration for ferromagnetic nearest-neighbor and antiferromagnetic next-nearest-neighbor exchange interactions [30]. For double-perovskite systems, on the other hand, the number of different exchange paths is increased due to the monoclinic distortion. There are nine different exchange paths, three Mn–O–Ni bonds ($J_1$, $J_2$, and $J_3$), three Mn–Mn bonds ($J_4$, $J_5$, and $J_6$), and three Ni–Ni bonds ($J_7$, $J_8$, and $J_9$), as shown in Fig. 9. The sign of Mn–O–Ni superexchange interactions in INMO has been unknown, though the interaction between the half-filled $e_g$ orbital of $Ni^{2+}$ and the vacant $e_g$ orbital of $Mn^{4+}$ through the $p$ orbital of $O^{2-}$ is ferromagnetic for $180°$ bond, according to the Kanamori-Goodenough rule [31,32]. Some theoretical works for $R_2NiMnO_6$ have predicted that the Mn–O–Ni interaction changes from ferromagnetic to antiferromagnetic with decreasing ionic radius of the $A$ site, and up-up-down-down antiferromagnetic structure in $Y_2NiMnO_6$ [15] and INMO [20,21]. However, these theoretical results are not in agreement with the experimental results of ferromagnetic $Y_2NiMnO_6$ [19] and incommensurate INMO determined by the present study. We suggest here that the strong competition among those nine neighboring exchange interactions plays an important role for emergence of the incommensurate magnetic ordering in INMO. For further understanding the appearance of this magnetic ground state in INMO, additional theoretical calculations taking account of all possible exchange paths are desirable.

## V. CONCLUSION

A ferroelectric polarization $P \simeq 30$ $\mu C/m^2$ was observed in polycrystalline sample of the double-perovskite INMO below $T_N = 26$ K. The existence of the polarization is associated with the complex noncollinear magnetic ordering with an incommensurate propagation vector, $k = (0.274, 0, -0.0893)$ at 1.5 K, resulting in the polar magnetic point group $21'$. The noncollinear magnetic structure with "cycloid" and "proper screw" components and the $k$ vector with two-dimensional modulation is unique in perovskite systems. We suggested that the competition among many neighboring exchange interactions characteristic of the monoclinic crystal plays an important role for onset of such a unique magnetic structure. The induced macroscopic polarization can be understood in terms of the inverse Dzaloshinskii-Moriya effect with two components: one is for the cycloid case $p_1 \propto r_{ij} \times (S_i \times S_j)_\perp$ and the other one is $p_2 \propto [r_{ij} \cdot (S_i \times S_j)_\parallel]A$, where the proper screw spin component is coupled to the "ferriaxial" crystal rotation characteristic of the $B$-site ordered perovskite system.

## ACKNOWLEDGMENTS

The images shown in Fig. 1 were depicted using the software VESTA [33] developed by K. Momma. The work at ISIS was supported by TUMOCS project. This project has received funding from the European Union's Horizon 2020 research and innovation programme under the Marie Skłodowska-Curie Grant Agreement No. 645660.